\documentstyle[amssymb,epsf,twocolumn,prb,aps,psfig]{revtex}

\begin{document}
\twocolumn[\hsize\textwidth\columnwidth\hsize\csname @twocolumnfalse\endcsname
\title{Common energy scale for magnetism and superconductivity in underdoped cuprates: a $\mu $SR investigation of (Ca$_{x}$La$_{1-x}$)(Ba$_{1.75-x}$La$_{0.25+x}$)Cu$_{3}$O$_{y}$}

\author{$^{1}$Amit Kanigel, $^{1}$Amit Keren, $^{1}$Yaakov Eckstein, $^{1}$Arkady
Knizhnik, $^{2}$James S. Lord, and $^{3}$Alex Amato}
\address{$^{1}$Physics Department, Technion-Israel Institute of Technology, Haifa\\
32000, Israel.\\
$^{2}$Rutherford Appleton Laboratory, Chilton Didcot, Oxfordshire OX11 0QX,\\
U.K.\\
$^{3}$Paul Scherrer Institute, CH 5232 Villigen PSI, Switzerland}
\date{\today}
\maketitle

\begin{abstract}
We characterize the spontaneous magnetic field, and determine the associated
temperature $T_{g}$, in the superconducting state of (Ca$_{x}$La$_{1-x}$)(Ba$%
_{1.75-x}$La$_{0.25+x}$)Cu$_{3}$O$_{y}$ using zero and longitudinal field $%
\mu $SR measurements for various values of $x$ and $y$. Our major findings
are: (I) $T_{g}$ and $T_{c}$ are controlled by the same energy scale, (II)
the phase separation between hole poor and hole rich regions is a {\em %
microscopic} one, and (III) spontaneous magnetic fields appear gradually
with no moment size evolution.
\end{abstract}

\pacs{PACS numbers: 74.0}


]
There is growing evidence that at low temperatures (T), cuprates
phase-separate into regions that are hole ``poor'' and hole ``rich'' \cite
{Kapitulnik1}. While hole rich regions become superconducting below $T_{c}$,
the behavior of hole poor regions at these temperatures is not quite clear.
Some data support the existence of magnetic moments in these regions. In
impure cases, like Zn or Li doped YBCO, the impurity creates both the hole
poor regions \cite{NachumiPRL96} and the magnetic moments \cite
{MendelsPRB94,Bobroff2}. In pure cases, such as LSCO, these magnetic moments
are created spontaneously and undergo a spin glass like freezing at $T_{g}$ 
\cite{Nieder1}. However, there are still many open questions regarding these
moments and the spontaneous magnetic fields associated with them. For
example: Is there a true phase transition at $T_{g}$? What is the field
profile and how is it different from, or similar to, a canonical spin glass?
Is the field confined solely to the hole poor regions or does it penetrate
the hole rich regions? Also, the interplay between magnetism and
superconductivity is not clear. Is strong magnetic background beneficial or
detrimental to superconductivity?

\begin{figure}[tbp]
\centerline{\epsfxsize=7.5cm \epsfbox{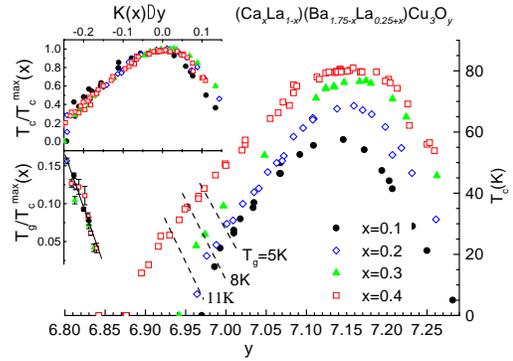}}
\caption{Phase diagram of (Ca$_{x}$La$_{1-x}$)(Ba$_{1.75-x}$La$_{0.25+x}$)Cu$%
_{3}$O$_{y}$. The dashed lines indicate samples with equal $T_{g}$. Insets: $%
T_{c}/T_{c}^{\max }$ and $T_{g}/T_{c}^{\max }$ as a function of $K(x)\Delta y
$ where $\Delta y=y-7.15$, and $K(x)$ is chosen so that all $%
T_{c}/T_{c}^{\max }$ data sets collapse to a single curve.}
\label{clblco}
\end{figure}

We address these questions by performing zero and longitudinal field muon
spin relaxation experiments on a series of polycrystalline (Ca$_{x}$La$%
_{1-x} $)(Ba$_{1.75-x}$La$_{0.25+x}$)Cu$_{3}$O$_{y}$ (CLBLCO) samples. This
superconductor belongs to the 1:2:3 family and has several properties that
make it ideal for our purpose. It is tetragonal throughout its range of
existence $0\leq x\lessapprox 0.5$, so there is no ordering of CuO chains.
Simple valence sums \cite{comment3}, more sophisticated bond-valance
calculations \cite{Eckstein1}, and thermoelectric power measurements \cite
{Arkady1} show that the hole concentration is $x$ independent. As shown in
Fig.~\ref{clblco}, by changing $y$, for a constant value of $x$, the full
superconductivity curve, from the under-doped to the over-doped, can be
obtained. Finally, for different Ca contents, parallel curves of $T_{c}$ vs $%
y$ are generated. Therefore, with CLBLCO one can move continuously, and with
minimal structural changes, from a superconductor resembling YBCO to one
similar to LSCO.

The preparation of the samples is described elsewhere \cite{clblco1}. Oxygen
content was determined using iodometric titration. All the samples were
characterized using X-ray diffraction and were found to be single phase. $%
T_{c}$ is obtained from resistivity measurements. We also verified using TF-$%
\mu $SR that CLBLCO respects the Uemura relations \cite{TheAmits2} and that
it is a bulk superconductor.

The $\mu $SR experiments were done at two facilities. When a good
determination of the base line was needed we used the ISIS pulsed muon
facility, Rutherford Appleton Laboratory, UK. When high timing resolution
was required we worked at Paul Scherrer Institute, Switzerland (PSI). Most
of the data were taken with a $^{4}$He cryostat. However, in order to study
the internal field profile we had to avoid dynamical fluctuations by
freezing the moments completely. For this purpose we used the $^{3}$He
cryostat at ISIS with a base temperature of $350$ mK.

\begin{figure}[tbp]
\centerline{\epsfxsize=7.5cm \epsfbox{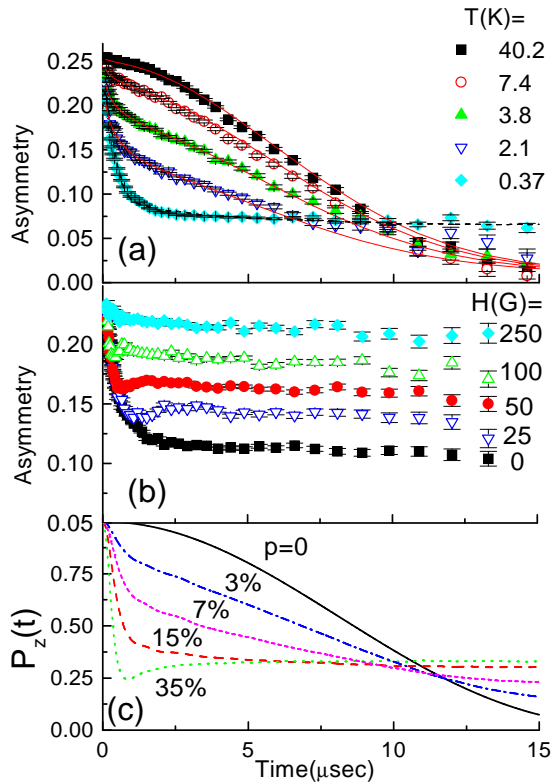}}
\caption{(a) ZF-$\protect\mu $SR spectra obtained in a $x=0.1$, $y=7.012$
sample. The solid lines are fit to the data using Eq.~\ref{func}, the dashed
line is a fit using the simulation as described in the text. (b) $\protect%
\mu $SR spectra obtained in longitudinal fields from the $x=0.4$, $y=6.984$
sample at 350mK. (c) Polarization curves generated by the simulation program
as described in the text.}
\label{pol}
\end{figure}

Typical muon asymmetry [$A(t)$] depolarization curves, proportional to the
muon polarization $P_{z}(t)$, are shown in Fig.~\ref{pol} (a) for different
temperatures in the $x=0.1$ and $y=7.012$ ($T_{c}=33.1K$) sample. The change
of the polarization shape with temperature indicates a freezing process, and
the data can be divided into three temperature regions. In region (I), given
by $T\gtrsim 8$~K, the muon relaxes according to the well known Kubo-Toyabe
(KT) function, typical of the case where only frozen nuclear moments are
present\cite{MusrBook}. In region (II), bounded by $8$ K $\gtrsim T\gtrsim 3$%
~K, part of the polarization relaxes fast and the rest as in the first
region. As the temperature is lowered the fast portion increases at the
expense of the slow one. Moreover, the relaxation rate in the fast portion
seems independent of temperature. Finally, at long time the asymmetry
relaxes to zero. In region (III), where $3$ K $\gtrsim T$, the asymmetry at
long times no longer relaxes to zero, but instead recovers to a finite
value. This value is $\simeq 1/3$ of the initial asymmetry $A_{z}(0)$.

To demonstrate that the internal field is static at base temperature, the
muon polarization was measured with an external field applied parallel to
the initial muon spin-polarization. This geometry allows one to distinguish
between dynamic and static internal fields. In the dynamic case the
asymmetry is field independent \cite{Comment1}. In contrast, in the static
case the total field experienced by the muon is a vector sum of $H$ and the
internal fields, which are of order $\left\langle B^{2}\right\rangle ^{1/2}$%
. For $H\gg \left\langle B^{2}\right\rangle ^{1/2}$ the total field is
nearly parallel to the polarization. Therefore, in the static case, as $H$
increases, the depolarization decreases, and the asymmetry recovers to its
initial value. Because we are dealing with a superconductor this field sweep
was done in field-cool conditions. Every time the field was changed the
sample was warmed up above $T_{c}$ and cooled down in a new field. The
results are shown in Fig.~\ref{pol}(b). At an external field of 250~G, the
total asymmetry is nearly recovered. Considering the fact that the internal
field is smaller than the external one due to the Meissner effect, this
recovery indicates that the internal field is static and of the order of
tens of Gauss.

We divide the data analysis into two parts: high temperatures (region II),
and base temperature. First we discuss region II. Here we focus on the
determination of $T_{g}$. For that purpose we fit a combination of a fast
relaxing function and a KT function to the data \cite{Savici}
\begin{equation}
A_{z}(t)=A_{m}\exp \left( -\sqrt{\lambda t}\right) +A_{n}KT(t),  \label{func}
\end{equation}
where $A_{m}$ denotes the amplitude of the magnetic part, $\lambda $ is the
relaxation rate of the magnetic part, and $A_{n}$ is the amplitude of the
nuclear part. The relaxation rate of the KT part was determined at high
temperatures and is assumed to be temperature independent. The sum $%
A_{m}+A_{n}$ is constrained to be equal to the total initial asymmetry at
high temperatures. The relaxation rate $\lambda $ is common to all
temperatures. The solid lines in Fig.~\ref{pol} are the fits to the data
using Eq.~\ref{func}.

\begin{figure}[tbp]
\centerline{\epsfxsize=7.5cm \epsfbox{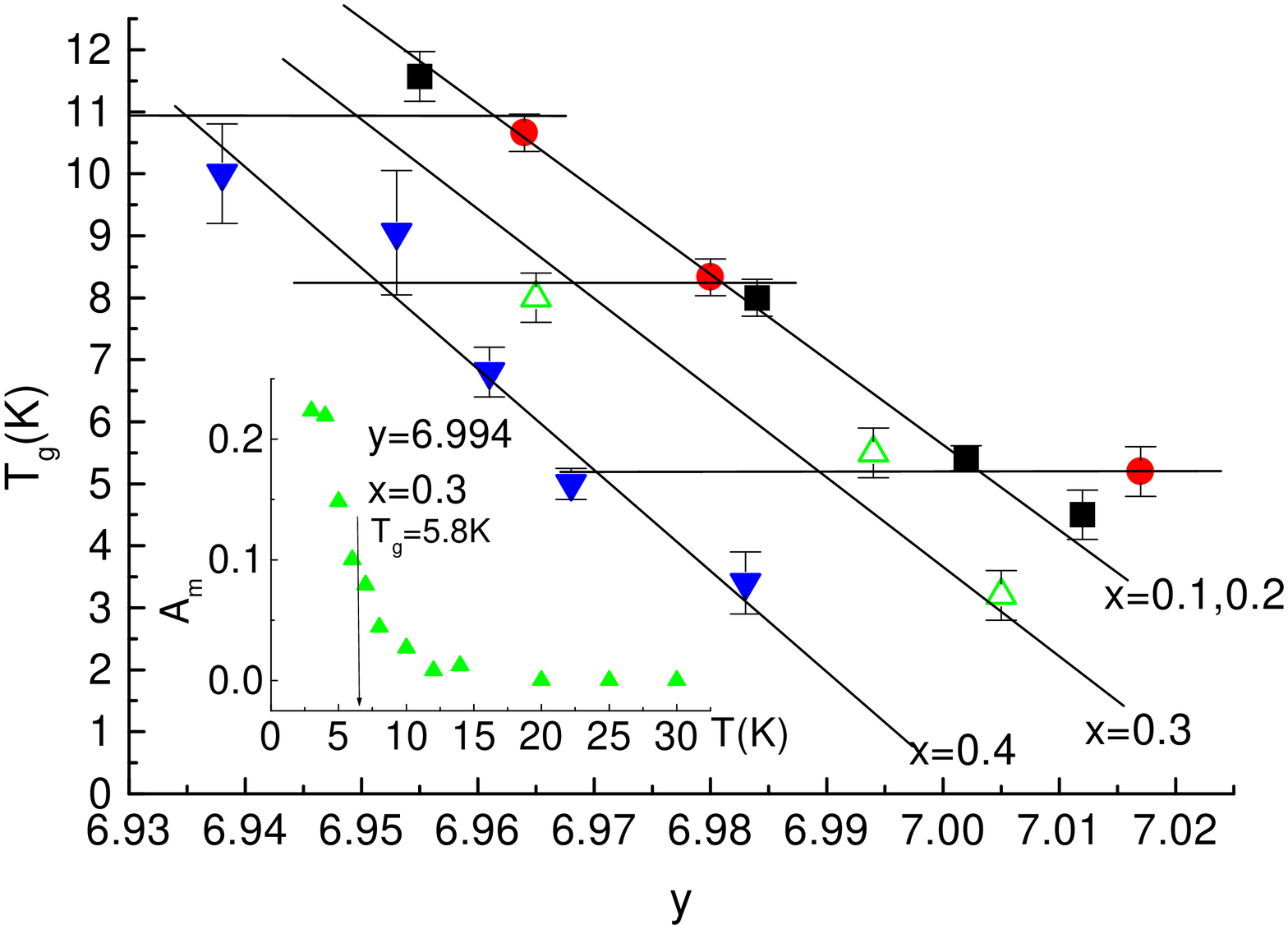}}
\caption{$T_{g}$ vs. $y$. The horizontal solid lines are the equal $T_{g}$
lines appearing in Fig \ref{clblco}. Inset: Magnetic amplitude as function
of temperature for a $x=0.3$ $y=6.994$ sample. The arrow indicates $T_{g}$
of that sample.}
\label{Tg}
\end{figure}

The success of this fit indicates the simultaneous presence of two phases in
the sample; part of the muons probe the magnetic phase while others probe
only nuclear moments. As the temperature decreases $A_{m}$, which is
presented in the insert of Fig.~\ref{Tg}, grows at the expense of $A_{n}$.
At low temperatures $A_{m}$ saturates to the full muon assymettry. A similar
temperature dependecne of $A_{m}$ is found in all samples. The origin of the
magnetic phase is electronic moments that slow down and freeze in a random
orientation. The fact that $\lambda $ is temperature independent means that
in the magnetic phase $\gamma _{\mu }\left\langle B^{2}\right\rangle ^{1/2}$%
, where $\gamma _{\mu }$ is the muon gyromagnetic ratio, is temperature
independent. In other words, as the temperature is lowered, more and more
parts of the sample become magnetic, but the moments in these parts saturate
upon freezing.

Our criterion for $T_{g}$ is the temperature at which $A_{m}$ is half of the
total muon polarization as demonstrated by the vertical line in the insert
of Fig.~\ref{Tg}. The phase diagram that is shown in Fig.~\ref{Tg}
represents $T_{g}$ for various samples differing in Ca and O contents. This
diagram is systematic and rather smooth suggesting good control of sample
preparation. As expected, for constant $x$, higher doping gives lower $T_{g}$%
.

We have singled out three groups of samples with a common $T_{g}=11$, $8$
and $5$~K as shown in Fig.~\ref{Tg} by the horizontal solid lines. These
samples are represented in the phase diagram in Fig.~\ref{clblco} by the
dotted lines. The phase diagram, containing both $T_{g}$ and $T_{c}$, is the
first main finding of this work. It provides clear evidence of the important
role of the magnetic interactions in high temperature superconductivity. In
fact, this phase diagram is consistent with recent theories \cite{Assa} of
hole pair boson motion in an antiferromagnetic background. Those theories
conclude that $T_{c}\propto Jn_{s}$ where $n_{s}$ is the superconducting
carrier density, and $J$ is the antiferromagnetic coupling energy
 \cite{UemuraNature}. From the measurements at constant $x$ we
 see that $T_{g}\propto Jf(n_{s})$
where $f$ is some decreasing function of $n_{s}$. We assume that $%
n_{s}=n_{s}[K(x)\Delta y]$ where $\Delta y=y-7$.$15$ is chemical doping
measured from optimum, and $K$ is a scaling parameter which relates chemical
to mobile-charge doping. Since $T_{c}^{\max} \propto J n_s($optimum$)$, both 
$T_{c}/T_{c}^{\max }$ and $%
T_{g}/T_{c}^{\max }$ should be functions only of $K(x)\Delta y$. We find $K(x)$
by making all $T_{c}/T_{c}^{\max }$ collapse onto one curve. This is
demonstrated in the upper inset of Fig.~\ref{clblco}. Using these values of $%
K(x)$ we also plot $T_{g}/T_{c}^{\max }$ as a function of $K(x)\Delta y$ in the
lower inset of Fig.~\ref{clblco}. Again all data sets
collapse onto a single curve. This indicates that the same single energy 
scale $J$
controls both the superconducting and magnetic transitions.

\begin{figure}[tbp]
\centerline{\epsfxsize=7.5cm \epsfbox{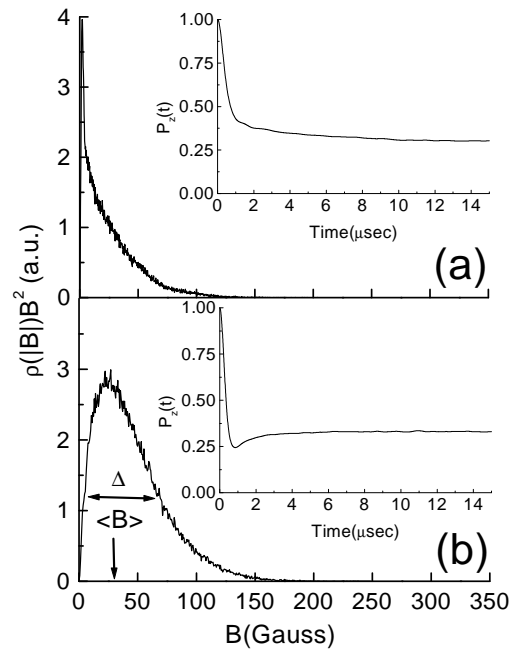}}
\caption{(a) The internal field distribution extracted from the simulations
for the case of correlation length $\protect\xi =3$ lattice constants,
maximum moment size of $0.06\protect\mu_B$ and magnetic moment concentration 
$p=15$\%. Inset: The muon spin polarization for that distribution. (b) The
same as above for the case of $p$=35\%.}
\label{model}
\end{figure}

We now turn to discuss the muon depolarization at base temperature. In this
case all the muons experience only a static magnetic field, as proven above.
This allows one to reconstruct the internal field distribution out of the
polarization curve. The polarization of a muon spin experiencing a unique
field ${\bf B}$ is given by $P_{z}(t)=\cos ^{2}(\theta )+\sin ^{2}(\theta
)\cos (\gamma |{\bf B}|t)$, where $\theta $ is the angle between the field
and the initial spin direction. When there is an isotropic distribution of
fields, a 3D powder averaging leads to 
\begin{equation}
P_{z}(t)=\frac{1}{3}+\frac{2}{3}\int_{0}^{\infty }\rho (|B|)\cos (\gamma
|B|t)B^{2}dB
\end{equation}
where $\rho (|B|)$ is the distribution of $\left| {\bf B}\right| $.
Therefore, the polarization is given by the Fourier transform of $\rho
(|B|)B^{2}$ and has a $1/3$ base line. When the distribution of ${\bf B}$ is
centered around zero field, $\rho (|B|)B^{2}$ is a function with a peak at $%
\langle B\rangle $ and a width $\Delta $, and both these numbers are of the
same order of magnitude [e.g. Fig~\ref{model}(b)]. Therefore we expect the
polarization to have a damped oscillation and to recover to $1/3$, a
phenomenon known as the dip [e.g. the inset in Fig~\ref{model} (b)].
Gaussian, Lorentzian and even exponential random field distribution \cite
{Larkin}, and, more importantly, all known canonical spin glasses, produce
polarization curves that have a dip before the $1/3$ recovery. Furthermore,
a dipless polarization curve that staruates to $1/3$ cannot be explanied
using dynamical arguments. Therefore, the most outstanding feature of the
muon polarization curve at base temperature is the fact that no dip is
present, although there is a $1/3$ tail. This behavior was found in all of
our samples with $T_{c}>7$~K, and also in Ca doped YBCO \cite{Nieder2} and
Li doped YBCO \cite{Mendels_Private}.

The lack of the dip in $P_{z}(t)$ can tell much about the internal field
distribution. It means that $\langle B\rangle $ is much smaller than $\Delta 
$. In that case the oscillations will be over-damped and the polarization
dipless! In Fig.~\ref{model} we show, in addition to the $\langle B\rangle
\simeq \Delta $ case described above [panel (b)], a field distribution that
peaks around zero [panel (a)]. Here $\langle B\rangle $ is smaller than $%
\Delta $, and, indeed, the associated polarization in the inset is dipless.
Thus in order to fit the base temperature polarization curve we should look
for $\rho (|B|)B^{2}$ with most of its weight around zero field. This means
that $\rho (|B|)$ diverges like $1/B^{2}$ at $|B|\rightarrow 0$, namely,
there is abnormally high number of low field sites.

It also means that the phase separation is not a macroscopic one. If it
were, all muons in the field free part would probe only nuclear moment and
their polarization curve should have a dip or at least its begining as in
the high temperature data. The same would apply for the total polarization
curve, in contrast to observation. Thus, the superconducting and magnetic
regions are intercalated on a microscopic scale ($\sim 20\AA $) \cite
{comment2}. This is the second main finding of this work.

The special internal field distribution, and the nature of the gradual
freezing of the spins, can be explained by the intrinsic inhomogeneity of
hole concentration. The part of the sample that is hole poor, and for that
reason is ``more'' antiferromagnetic, will freeze, while the part which is
hole rich will not freeze at all. The variation in the freezing temperature
of different parts of the sample can be explained by the distribution of
sizes and hole concentration in these antiferromagnetic islands \cite{Cho}.
The large number of low field sites is a result of the fact that the
magnetic field generated in the magnetic regions will penetrate into the
hole rich regions but not completely.

To improve our understanding of the muon polarization, we performed
simulations of a toy model aimed at reproducing the results described above.
A 2D $100\times 100$ square lattice is filled with two kinds of moments,
nuclear and electronic. All the nuclear moments are of the same size, they
are frozen and they point in random directions. Out of the electronic
moments only a small fraction $p$ is assumed to be frozen; they represent
magnetic regions with uncompensated antiferromagnetic interactions. Since
these regions may vary in size the moments representing them are random, up
to a maximum size. The frozen electronic moments induce spin polarization in
the other electronic moments surrounding them. Following the work of others 
\cite{Bobroff1}, we use decaying staggered spin susceptibility which we take
to be exponential, namely, 
\begin{equation}
\chi ^{^{\prime }}({\bf r})=(-1)^{n_{x}+n_{y}}\exp (-r/\xi )
\end{equation}
where ${\bf r}=n_{x}a\widehat{x}+n_{y}a\widehat{y}$ represents the position
of the neighbor Cu sites, ${\bf a}$ is the lattice vector, and $\xi $ is the
characteristic length scale. Because of this decay, at low frozen spin
concentration, large parts of the lattice are practically field free (expect
for nuclear moments). However, the important point is that no clear
distinction between magnetic and field free (superconducting) regions exists.

The muon polarization time evolution in this kind of field distribution is
numerically simulated. The interaction between the muon and all the other
moments is taken to be dipolar, and $\xi $ is taken to be 3 lattice
constants \cite{Kapitulnik1,NachumiPRL96}. The dashed line in Fig.~\ref{pol}
is a fit to the $T=350$ mK data, which yield $p=15$\% and maximum moment
size $\simeq 0.06\mu _{B}$ . As can be seen, the line fits the data very
well. However, as expected, the fit is sensitive to $p \xi^2 $ only, namely
the effective area of the magnetic islands, so longer $\xi $ would have
given smaller $p$. The field distributions and the polarization curve shown
in Fig.~\ref{model} were actually generated using the simulation. In (a) the
spin density is 15\% while in (b) the density is 35\%.

In panel (c) of Fig.~\ref{pol} we show the spin polarization for different
hole concentration, varying from 0\% to 35\% with the same $\xi =3$. The
resemblance between the simulation results and the muon polarization as a
function of temperature in panel (a) leads us to our third conclusion that
the freezing process is mostly a growth in the total area of the frozen AF
islands.

We are now in a position to address the questions presented in the
introduction. The appearance of spontaneous magnetic field in CLBLCO is a
gradual process. As the temperature is lowered microscopic regions of frozen
moments appear in the samples, and their area increases but the moments do
not. In the ground state the field profile is very different from that of a
canonical spin glass or any other standard magnet. It could only be
generated by microscopic intercalation of an abnormal number of zero field
regions with magnetic regions without a clear distinction between the two.
Finally, and most importantly, the phase diagram containing both $T_{c}$ and 
$T_{g}$ leads us to believe that these temperatures are determined by the
same energy scale given by $J$.

We would like to thank the PSI and ISIS facilities for their kind
hospitality and continuing support of this project. We acknowledge very
helpful discussions with Assa Auerbach and Ehud Altman. This work was funded
by the Israeli Science Foundation, the EU-TMR program, and the Technion V.
P. R fund - Posnansky, and P. and E. Nathen, research funds.


\begin{references}
\bibitem{comment3}  The density of hole per plane is $p=-0.205+y/3$.
\bibitem{Comment1}  As long as $\gamma _{\mu }H<\nu $, where $\nu $ is the
fluctuation rate.
\bibitem{comment2}  A moment of $1\mu_{B}$ will induce a 1G field at a
distance of $\sim20 \AA$. This field is equivalent to the nuclear moments
background.
\end{references}
\end{document}